\documentclass[letterpaper,twocolumn,10pt]{article}
\usepackage{usenix,epsfig,endnotes}
\usepackage{multirow}
\usepackage{booktabs}
\usepackage{array}
\usepackage{graphicx} 
\usepackage{epstopdf}
\usepackage[ruled]{algorithm2e}
\usepackage{url}

\begin{document}
	
	\date{}
	
	\title{\Large \bf ICSTrace: A Malicious IP Traceback Model\\for Attacking Data of Industrial Control System}
	
	\author{
		{\rm Feng Xiao}\\
		Anhui Province Key Laboratory of\\
		Big Data Analysis and Application,\\
		School of Computer Science and Technology,\\
		University of Science and Technology of China\\
		xiaof686@mail.ustc.edu.cn
		\and
		{\rm Qiang Xu}\\
		Electronic Engineering Institute of\\
		Hefei, China\\
		yfnm126@126.com
	} 
	
	\maketitle
	
	\thispagestyle{empty}

	\subsection*{Abstract}
	Considering the attacks against industrial control system are mostly organized and premeditated actions, IP traceback is significant for the security of industrial control system. Based on the infrastructure of the Internet, we have developed a novel malicious IP traceback model-ICSTrace, without deploying any new services. The model extracts the function codes and their parameters from the attack data according to the format of  industrial control protocol, and employs a short sequence probability method to transform the function codes and their parameter into a vector, which characterizes the attack pattern of malicious IP addresses. Furthermore, a Partial Seeded K-Means algorithm is proposed for the pattern's clustering, which helps in tracing the attacks back to an organization. ICSTrace is evaluated basing on the attack data captured by the large-scale deployed honeypots for industrial control system, and the results demonstrate that ICSTrace is effective on malicious IP traceback in industrial control system.
	
	\section{Introduction}
	With the rapid development of the Internet of Things (IoT), more and more Industrial Control Systems (ICS) are connected into the Internet. As the key bond between the virtual signal and the real equipment, an Internet-connected ICS makes the production process to be more accurate and agile. But it also narrows the distance between the cyber attacks and the industrial infrastructure. As we know, Stuxnet worm was disclosed to be the first worm attacking the energy infrastructure \cite{chen2011lessons,kushner2013real} in 2010. In 2014 the hackers attacked a steel plant in Germany, so that the blast furnace can not be closed properly \cite{zetter2015cyberattack}. On December 23, 2015, the Ukrainian power network suffered a hacker attack, which was the first successful attack to the power grid, resulting in hundreds of thousands of users suffering power blackout for hours \cite{zetter2016inside}. In 2017, the security vendor ESET disclosed an industrial control network attack weapons named as win32/Industroyer, which implemented malicious attacks on power substation system \cite{eseturl}.\
	
	ICSs are highly interconnected and interdependent with the critical national infrastructure \cite{stouffer2011guide}, and thus the attackers have noticed the high returns to attack ICS in recent years. The attackers are diverse in identity. They may be hackers, members of organized criminal groups, or even a hostile country. The worse situation is that ICS has become the new target of terrorists to gain the influence by destroying the real physical world. As traditional ICS is physically isolated from the Internet, most researches just focus on the functional safety of the system rather than the security consideration of the network. There are not any special protective measures, not to mention the attribution mechanism for tracing the attack back \cite{knowles2015survey}. Security researchers are now committed to the intrusion detection technology for ICS. They want to identify, intercept and alert the threats, before a severe attack occurs. These intrusion detection technologies can be divided into several categories as follows: state-based \cite{khalili2015sysdetect}, behavior-based \cite{kwon2015behavior}, rule-based \cite{yang2013rule}, characteristic-based \cite{mcparland2014monitoring}, model-based \cite{mo2014detecting}, and ML-based (machine learning) \cite{shang2014modbus,zhou2015design}.\
	
	Because ICS plays an important role in the critical national infrastructure, the cyber attacks against ICS are mostly organized and premeditated actions. It is significant not only to determine whether there is a threat in ICS, but also to trace the attack back. Furthermore, locating the initiators and their motivations before or during an attack is crucial for deterring and cracking down the premeditated and organized attackers.\
	
	Attribution is one of the most intractable problems of an emerging field, created by the underlying technical architecture and geography of the Internet \cite{rid2015attributing}. The current dominant IP traceback technologies include packet marking mechanism \cite{savage2000practical}, packet logging mechanism \cite{snoeren2001hash} and their hybrid \cite{gong2005ip,gong2008more}. Packet marking mechanism needs the routers to write a tag (for example, IP address) into some fields of every packet. The target retrieves all the tags from the received packets and finds out the routing path. Packet marking mechanism includes two categories: Probabilistic Packet Marking (PPM) \cite{savage2000practical} and Deterministic Packet Marking (DPM) \cite{belenky2003ip}. Packet logging mechanism needs the routers to record all the forwarded packets so as to reveal the routing path. Apparently, this mechanism consumes a lot of storage space. All these IP traceback technologies above need to re-design the Internet or to deploy new services. There is still no applicable IP traceback system to deploy over the network.\
	
	The ultimate goal of attribution is identifying an organization or a government, not individuals \cite{rid2015attributing}. Our study identifies an organization by zooming down to a single IP level and then zooming back out to an organization or a unit level without changing the Internet architecture or deploying new services. Instead of tracing back to the source of a packet directly, we just recognize the malicious IP addresses which belong to the same organization.\
	
	In this study, we present a malicious IP traceback model (ICSTrace) for industrial control system, and this model makes the following contributions:\
	
	\begin{enumerate}
		\item Based on the deep analysis of ICS protocol S7comm, the function codes and their parameters are extracted from the attack data.
		\item A feature vector of the function codes and their parameters are designed to represent the attack patterns.
		\item The slide window method is adopted to reduce the dimension of those multidimensional samples.
		\item A Partial Seeded K-Means clustering algorithm is proposed based on K-Means algorithm.
		\item ICSTrace is proven to be effective basing on the real attack data captured by the large-scale deployed honeypots for ICS.
	\end{enumerate}\
	
	Section 2 introduces the research background and our previous work on the attack data collection. Section 3 gives the details of S7comm protocol. Section 4 describes the architecture of our IP traceback model. Section 5 and Section 6 introduce the attack pattern extraction method and Partial Seeded K-Means algorithm for clustering respectively. In Section 7, we evaluate our IP traceback model basing on the real attack data. Section 8 is our related works and Section 9 is the conclusion.
	
	\section{Background}
	
	ICS is a business process management and control system which is composed of various automatic control and process control components. It collects and monitors realtime signals to ensure the function of the automatic operation or the process control. Its application fields include program automation, industrial control, intelligent building, power transmission and distribution, smart meter, car communication and so on. ICS protocol refers to the communication protocol used in ICS. The most well-known ICS protocol includes S7, Modbus, BACnet, and DNP3.\
	
	At present, there is not any ICS attacking data set for security research. Therefore, we developed a high interactive ICS honeypot named as S7commTrace in previous work \cite{xiaofeng2017s7commtrace}, based on Siemens' S7comm protocol. Honeypot is a kind of security resource that is used to attract the attacker for illegal application without any business utility \cite{jicha2016scada}. Honeypot technology is a method to set some hosts, network services or information as a bait, to induce attackers, so that the behavior of the attacks can be captured and analyzed \cite{spitzner2003honeypots}. Honeypot can be used to better understand the landscape of where these attacks are originating \cite{zhuge2013honeypot}.\
	
	S7commTrace masquerades as a real PLC device by simulating the S7 protocol to capture the probing and attacking data. It can be divided into four modules, including TCP Communication module, S7comm Protocol Simulation module, Data Storage module and User Template, as shown in Figure 1.\
	
	\begin{figure}[t]
		\begin{center}
			\includegraphics[width=0.45\textwidth]{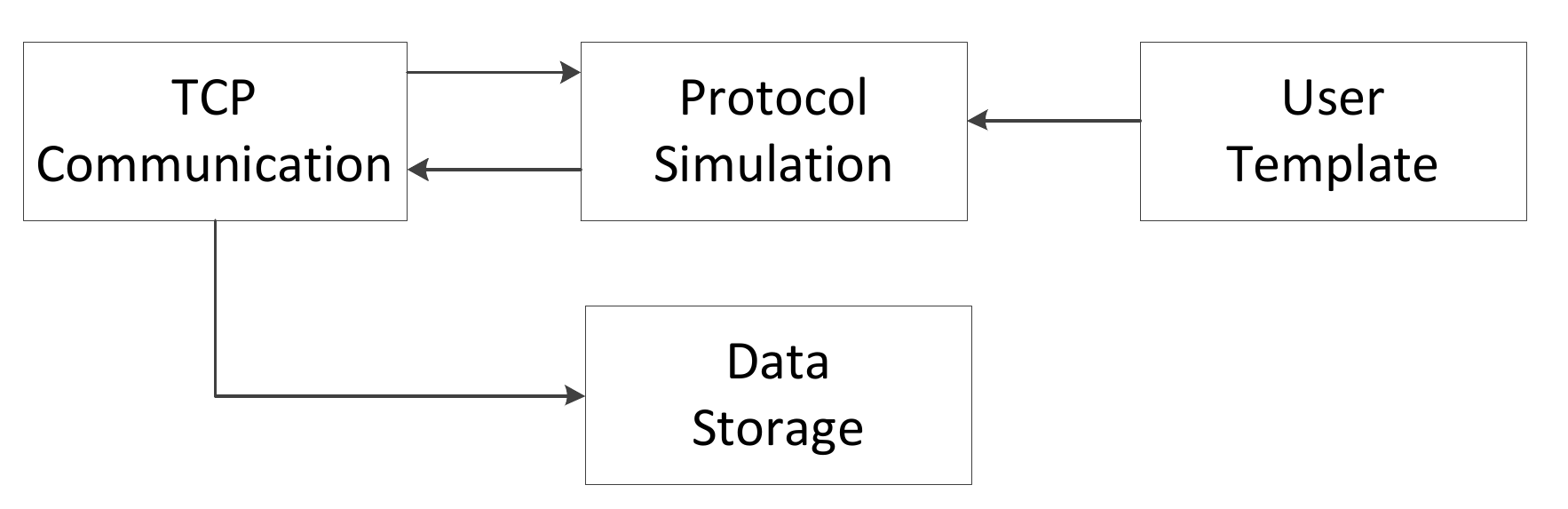}
		\end{center}
		\caption{Structure of ICS Honeypot (S7commTrace).}
	\end{figure}
	
	The main function of TCP Communication module is to listen on TCP port 102, submit the received data to the Protocol Simulation module, and reply to the remote peer. S7comm Protocol Simulation module parses the received data according to the protocol format and obtains the valid contents at first. And then S7comm Protocol Simulation module generates the reply data referring to User Template. At last, the reply data are sent back to TCP Communication module to be packaged. User Template records all the user-defined information such as PLC serial number, manufacturer, and so on. The Data Storage module handles the request and the response of data storage.\
	
	We deployed S7commTrace honeypots in United States, China, Germany, Russia, Japan, Singapore and Korea at the same time. The deployment utilize Aliyun and Host1Plus as virtual host with configuration of 1.5 Ghz single core CPU, 1GB RAM and 40GB Disk. All the operation systems of virtual hosts are Ubuntu Server. Every virtual host installs MySQL database to store data captured by S7commTrace.\
	
	Each S7commTrace ran for 272 days in average. At last, we captured 110501 requests of S7comm protocol, as shown in Table 1. In fact, not all requests are in accordance with S7comm format. Ignoring them, S7commTrace records a total of 46492 valid requests. If we define an uninterrupted TCP communication connection as a session, S7commTrace records 5797 sessions and 4224 valid sessions. Furthermore, a valid IP address indicates that this IP has at least one valid session.\
	
	\begin{table}
		\centering
		\begin{tabular}{p{3.5cm}<{\centering}p{3.3cm}<{\centering}}
			\toprule
			\textbf{Item} & \textbf{Count}\\
			\midrule
			Request	& 110501 \\
			Valid request & 46492 \\
			Session	& 5797 \\
			Valid session & 4224 \\
			IP address & 897 \\
			Valid IP address & 573\\
			\bottomrule
		\end{tabular}
		\caption{After 13 honeypots run for 272 days, count of all attack data and valid attack data.}
	\end{table}
	
	According to the DNS query results, we find that there are 26 IP addresses pointing to Shodan.io, 19 IP addresses pointing to eecs.umich.edu, 16 IP addresses pointing to neu.edu.cn, and 5 IP addresses pointing to plcscan.org, as shown in Table 2. This means 573 valid IP addresses belong to four organizations at least.\
	
	\begin{table}
		\begin{tabular}{p{2.3cm}<{\centering}p{2.3cm}<{\centering}p{2.0cm}<{\centering}}
			\toprule
			\textbf{Domain} & \textbf{Organization} & \textbf{IP number} \\
			\midrule
			Shodan.io & Shodan & 26 \\
			eecs.umich.edu & Censys & 19 \\
			neu.edu.cn & Ditecting & 16 \\
			plcscan.org & Beacon Lab & 5 \\
			Other & Unknown & 507 \\
			\bottomrule
		\end{tabular}
		\caption{IP statics by DNS reverse lookup.}
	\end{table}
	
	Shodan.io \cite{shodanurl} is the domain suffix of Shodan which is a search engine in cyberspace. In addition to retrieving traditional web services, Shodan has used the ICS protocol directly to crawl the ICS devices on the Internet, and visualizes their location and other information. Eecs.umich.edu is the domain suffix of the Department of Electrical and Computer Science (EECS) Department of University of Michigan, which is one of the agencies developing Censys \cite{censysurl,durumeric2015search}. Censys scans the devices in the Internet and stores the results in its database. It provides not only web and API query interfaces but also raw data to download. Neu.edu.cn is the domain suffix of Northeastern University of China which develops a search engine name as Ditecting \cite{ditectingurl}. Ditecting is capable of providing accurate information of ICS devices and their locations. Plcscan.org is the domain suffix of Beacon Lab \cite{beaconlaburl} which is committed to the research and the practice related to ICS security. These four organizations are the well-known security research institutes. They are scanning the devices in the Internet all the time, including the ICS devices. As shown in Table 2, except for the 66 IP addresses belonging to four well-known organizations, there are still 507 IP addresses which are resolved to be dynamic domain name or none domain name.
	
	\section{S7 Protocol}
	
	S7 protocol is a Siemens proprietary protocol \cite{s7commurl} running on programmable logic controllers (PLCs) of Siemens S7-200, 300, and 400series. It is suitable for either Ethernet, PROFIBUS or MPI networks . Because the objects of this study are those industrial control systems which are accessed to the Internet, we only discuss the TCP-based S7 protocol in Ethernet networks. As shown in Figure 2, S7 protocol packets are packed by COTP protocol, and then packed by TPKT protocol package for TCP connection.\
	
	\begin{figure}
		\centering
		\includegraphics[width=0.45\textwidth]{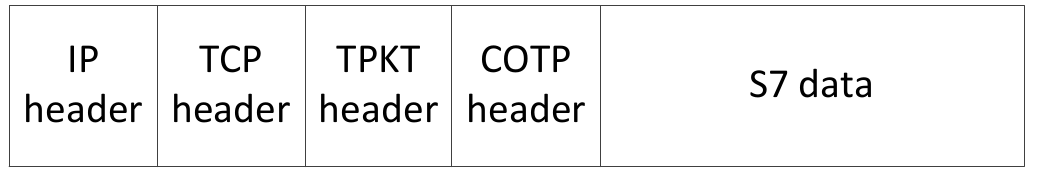}
		\caption{Header format of S7 communication packet.}
	\end{figure}
	
	As shown in Figure 3, the communication procedure of S7 protocol is divided into three stages. The first stage is to establish COTP connection, the second stage is to setup S7 communication, and the third stage is to exchange the request and the response for function code.\
	
	\begin{figure}[t]
		\begin{center}
			\includegraphics[width=0.45\textwidth]{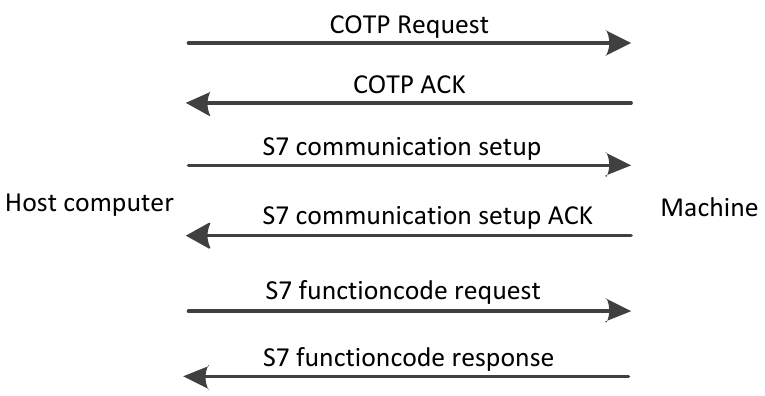}
		\end{center}
		\caption{Communication procedure of S7 protocol.}
	\end{figure}
	
	The Magic flag of the S7 protocol is fixed to 0x32, and the following fields are S7 type, data unit ref, parameters length, data length, result info, parameters and data, as shown in Figure 4.\
	
	\begin{figure*}[t]
		\begin{center}
			\includegraphics[width=0.9\textwidth]{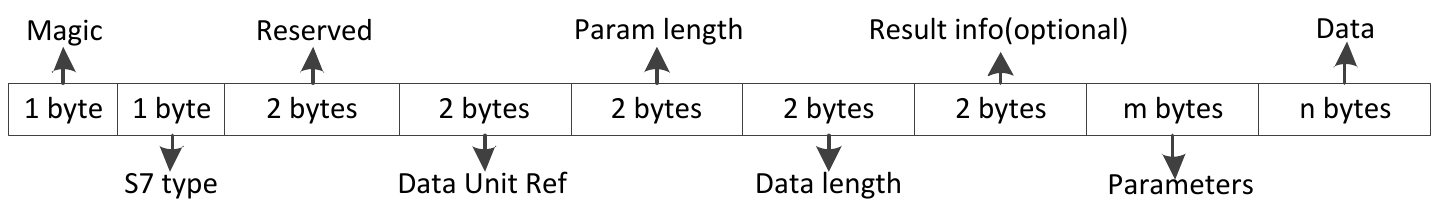}
		\end{center}
		\caption{Data format of S7 communication packet.}
	\end{figure*}
	
	In parameters field, the first byte stands for the function code of S7. Table 3 shows the optional function codes of S7. Communication Setup code is used to build a S7 connection; Read code helps the host computer to read data from PLC; Write code helps the host computer to write data to PLC.As for the codes of Request Download, Download Block, Download End, Download Start, Upload and Upload End, they are designed for downloading or uploading operations of blocks. PLC Control code covers the operations of Hot Run and Cool Run, while PLC Stop is used to turn off the device.\
	
	\begin{table}
		\begin{tabular}{p{2.5cm}<{\centering}p{4.5cm}<{\centering}}
			\toprule
			\textbf{Code} & \textbf{Function} \\
			\midrule
			0x00 & System Functions \\
			0x04 & Read \\
			0x05 & Write \\
			0x1a & Request Download \\
			0x1b & Download Block \\
			0x1c & Download End \\
			0x1d & Download Start \\
			0x1e & Upload \\
			0x1f & Upload End \\
			0x28 & PLC Control \\
			0x29 & PLC Stop \\
			0xf0 & Communication Setup \\
			\bottomrule
		\end{tabular}
		\caption{S7 protocol function code and the corresponding function.}
	\end{table}
	
	When the function code is 0x00, it stands for system function which is used to check system settings or status. And the details are described by the 4 bits function group code and 1 byte subfunciton code in the parameters field, as shown in Figure 5.\
	
	\begin{figure}[t]
		\begin{center}
			\includegraphics[width=0.45\textwidth]{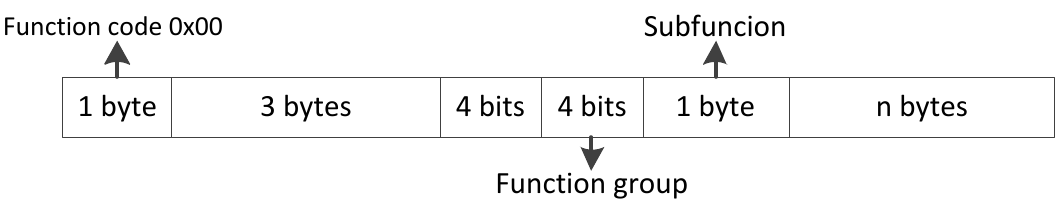}
		\end{center}
		\caption{When functioncode is 0x00, function group and subfunction position.}
	\end{figure}
	
	System Functions further divided into 7 groups, as shown in Table 4. Block function is used to read the block, and Time Function is used to check or set the device clock.\
	
	\begin{table*}
		\begin{tabular}{p{3.2cm}<{\centering}p{4.3cm}<{\centering}p{3.5cm}<{\centering}p{3.5cm}<{\centering}}
			\toprule
			\textbf{Function group code} & \textbf{Function} & \textbf{Subfunction code} & \textbf{Subfunction} \\
			\midrule
			\multirow{2}{*}{1} & \multirow{2}{*}{Programmer Commands} & 1 & Request diag data \\
			& & 2 & VarTab \\
			2 & Cyclic Data & 1 & Memory \\
			\multirow{3}{*}{3} & \multirow{3}{*}{Block Function} & 1 & List blocks \\
			& & 2 & List blocks of type \\
			& & 3 & Get block info \\
			\multirow{2}{*}{4} & \multirow{2}{*}{CPU Function} & 1 & Read SZL \\
			& & 2 & Message service \\
			5 & Security & 1 & PLC password \\
			6 & PBC BSEND/BRECV & None & None \\
			\multirow{3}{*}{7} & \multirow{3}{*}{Time Function} & 1 & Read clock \\
			& & 2,3 & Set clock \\
			& & 4 & Read clock (following) \\
			\bottomrule
		\end{tabular}
		\caption{When the function code is 0x00, it is system function and further divided into 7 groups.}
	\end{table*}
	
	\section{Structure of ICSTrace Model}
	
	When an attacker launches the attacks, he usually hides the IP address of his own resorting to the springboard host, VPN and other measures. As shown in Figure 6, after an ICS suffered an attack from the Internet, the security personnel can only see the last IP address connected to ICS instead of the real IP address of the attacker, not to mention the organization which belongs to.\
	
	\begin{figure}[t]
		\begin{center}
			\includegraphics[width=0.30\textwidth]{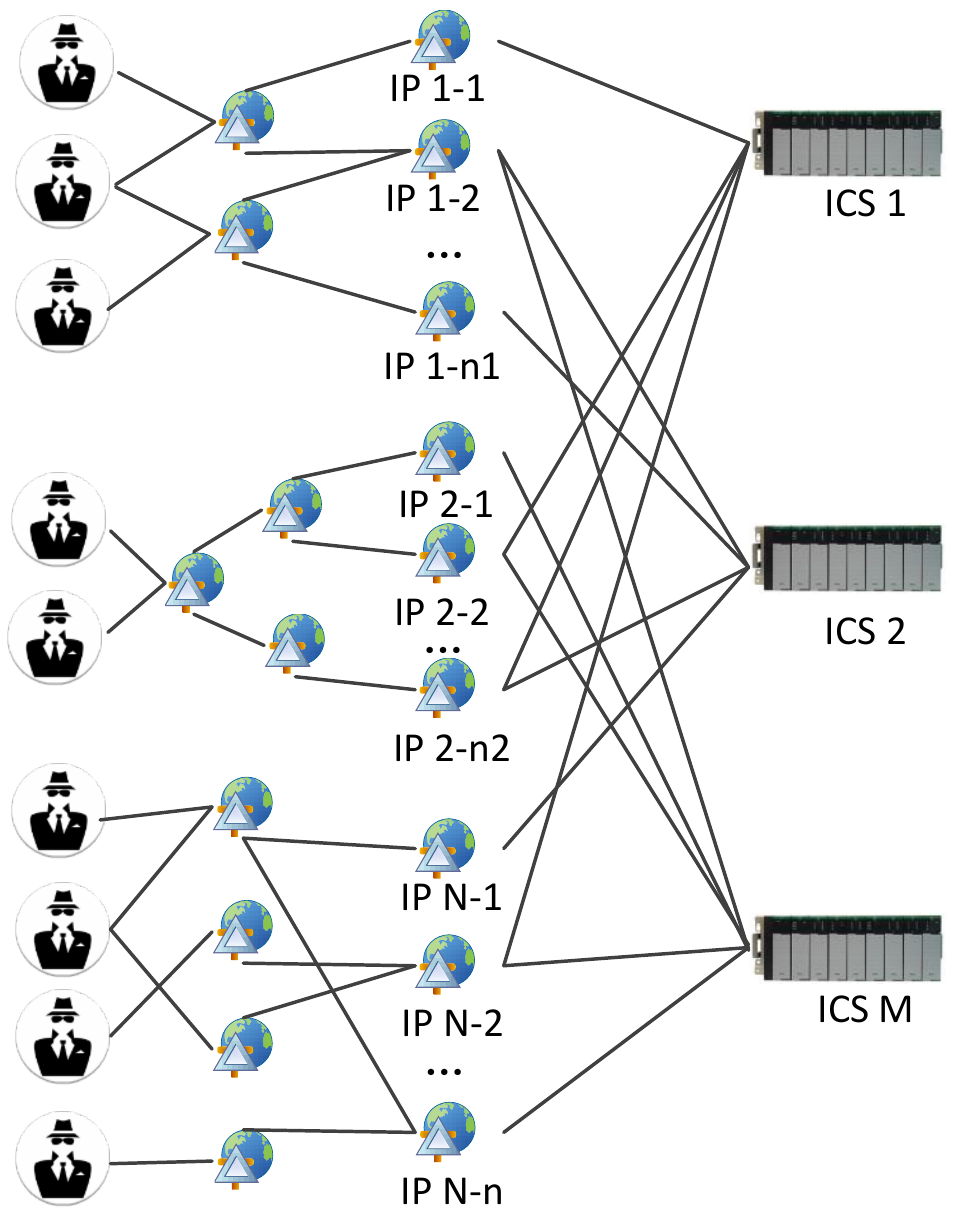}
		\end{center}
		\caption{Schematic diagram of attacking flow.}
	\end{figure}
	
	ICSTrace transforms the features of data from each IP address into a one-dimensional eigenvector. This eigenvector stands for the unique pattern of an attack. Therefore, the problem of attribution turns into a problem of clustering the patterns.\
	
	As shown in Figure 7, the input of ICSTrace is a malicious IP and its packets. The output is a cluster containing multiple IP addresses, which indicates an organization. ICSTrace model consists of three stages, including Protocol Resolution, Attack Pattern Extraction and Partial Seeded K-Means clustering. The main function of Protocol Resolution is to parse the packets and extract the function codes and their parameters. Attack Pattern Extraction transforms the function codes and their parameters into one-dimensional vector as the attack pattern of a certain IP address. Partial Seeded Means is used to cluster the attack patterns so that those IP addresses with the same patterns are aggregated into one cluster. And then, the cluster is labeled as a certain organization according to some auxiliary information (e.g. domain name or geographical location)of the IP addresses in it.\
	
	\begin{figure}[t]
		\begin{center}
			\includegraphics[width=0.45\textwidth]{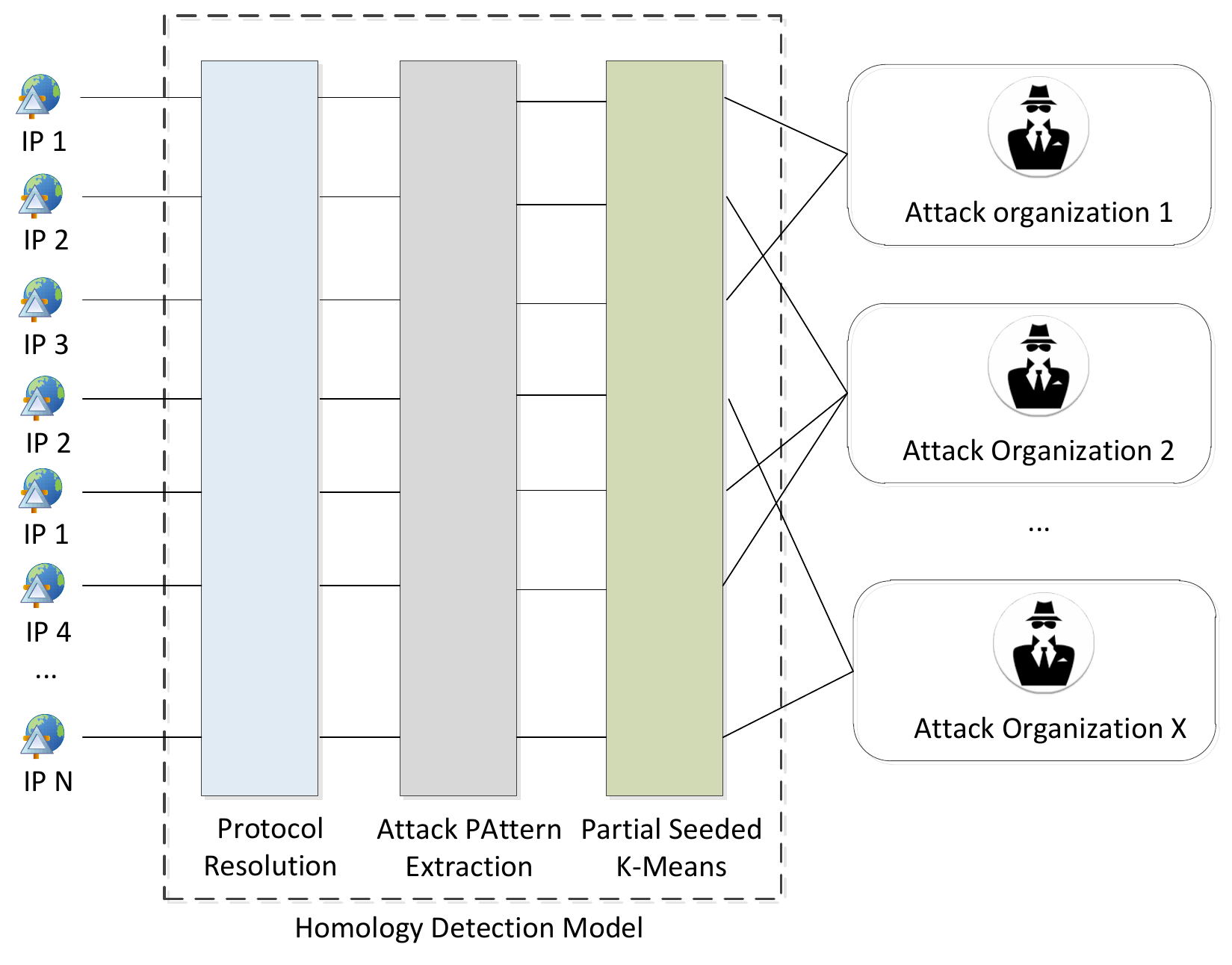}
		\end{center}
		\caption{Structure of ICS model.}
	\end{figure}
	
	\section{Attack Pattern Extraction}
	
	After an attacker has constructed the connection with ICS, he will carry out a series of delicate operations on purpose, which are expressed by the function codes and their parameters in table 3 and table 4. Therefore, the attacking features, which are extracted from the function codes and their parameters of S7comm protocol data, can reveal the intention of the attacker effectively.\
	
	As shown in Figure 8, one attacker may have several IP addresses to launch attacks. We have defined an uninterrupted TCP communication as a session, and one IP address may attack one or more ICSs for more than one times. And thus a single source IP may build several sessions. We call a packet sent by the attacker as a request and there are several packet interactions, so a session usually contains many requests.\
	
	\begin{figure}[t]
		\begin{center}
			\includegraphics[width=0.45\textwidth]{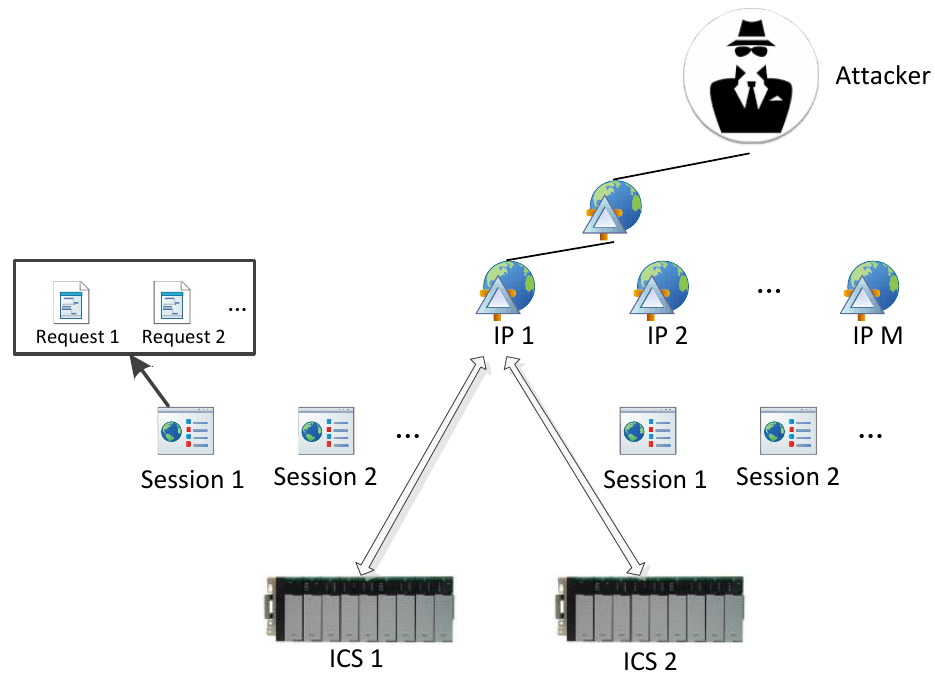}
		\end{center}
		\caption{Schematic diagram of attack-IP-session-request.}
	\end{figure}
	
	The function codes and their parameters of S7comm protocol are included in these requests, so we extract these from the communication data package, which is sent by the attacker to the receiver, as the feature of the attacker to construct IP traceback model.\
	
	\subsection{Mean Count of Function Codes and Parameters}
	Mean count of function codes (MCFC) refers to the average amount of the function codes of each session from the same IP address. Different attackers have different motivations, objectives and methods while conducting a cyber attack. As a result, quantities of requests and function codes are very different in different sessions.\
	$$MCFC=\frac{1}{n}\sum_{i=1}^{n}(Count\_of\_funciton\_codes)_{session_{i}},$$\
	$$session_{i}\in IP$$\
	
	Mean count of the parameters (MCP) refers to the average amount of the parameters used in the function codes of each session from the same IP address. Some function codes do not need parameters, some function codes need one or more parameters, so different attackers use different amount of parameters.\
	$$MCP=\frac{1}{n}\sum_{i=1}^{n}(Count\_of\_parameters)_{session_{i}},$$
	$$session_{i}\in IP$$
	\subsection{Function Codes Sequence and Parameters Sequence}
	the change rule of the function codes
	Function codes sequence (FCS) indicates the change rule of the function codes in all sessions from a single IP address. Different attackers may use the same kind of function codes while lunching an attack, but the chronological order is different. As shown in figure 9, the Function code $C_1,\,C_2,\,...,\,C_i$ can be arrayed to form a Markov chain in chronological order.\
	
	\begin{figure}[t]
		\begin{center}
			\includegraphics[width=0.45\textwidth]{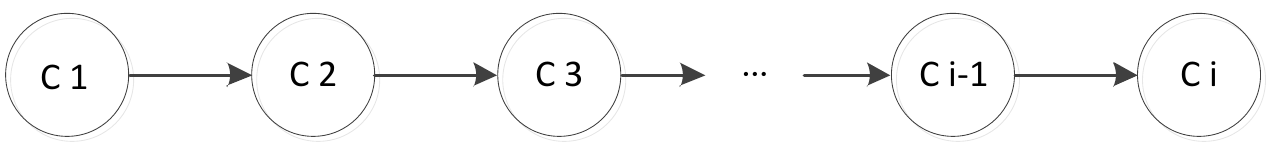}
		\end{center}
		\caption{Function codes sequence.}
	\end{figure}\
	
	Array the function codes in the session to form a function code sequence according to the chronological order.\
	$$F_{session_i}=(C_1,C_2,...,C_i),session_i\in IP$$\
	
	For some sessions may belong to the same source IP address, we combine the function codes serials and parameter serials of all sessions from the same IP address into a set of function code sequence.\
	$$
	F_n=\left(
	\begin{array}{c}
	F_{session_1}\\
	F_{session_2}\\
	...\\
	F_{session_n}
	\end{array}
	\right)=\left(
	\begin{array}{cccc}
	C_1 & C_2 & ... & C_{a_1} \\
	C_1 & C_2 & ... & C_{a_2} \\
	& & ... & \\
	C_1 & C_2 & ... & C_{a_n} \\
	\end{array}
	\right),$$
	$$session_{i}\in IP,1\leq i\leq n$$
	
	Different amount of sessions originate from each source IP and various methods are adopted by the attackers for each time, which results in the different function code sequences in each session. Therefore, $F_n$ of different source IP addresses are two-dimensional matrix vectors with unequal rows and columns.\
	
	These FCSs with uncertain amount and unequal length cannot be handled directly, for clustering algorithms like K-Means needs samples with same dimensions. In this study, we propose a method to convert these sequences with uncertain amount and unequal length into the vectors with same length, the detailed process is as follows:\
	
	Step 1 Add the start and the end status to the sequence.\
	
	For a sample set of sequence $F_n$, there are n sequences with unequal length and the length of which are $a_1,a_2, ...,a_n,\,a_i \geq 1,\,i \in [1,n]$ respectively. Add the start and the end status to each sequence in $F_n$, then we get ${F_n}'$. Now the length of each sequence is no less than 3.
	$$
	{F_n}'=\left(
	\begin{array}{cccccc}
	S & C_1 & C_2 & ... & C_{a_1} & E \\
	S & C_1 & C_2 & ... & C_{a_2} & E \\
	S & & & ... & & E \\
	S & C_1 & C_2 & ... & C_{a_n} & E \\
	\end{array}
	\right)
	$$
	
	Step 2 Get the unrepeatable set of short sequences.\
	
	Setting the window length equals 3 and the stride equals 1, we use the slide window method to process each sequence in ${F_n}'$. Then we get $a_1,a_2,...,a_n$ short sequences with the same length of 3, $a_i \geq 1,\,i \in [1,n]$. Then remove the duplicate sequences and add the short sequences into set $S=(s_1, s_2, ..., s_m), m \leq \sum_{i=1}^{n}a_i$.\
	
	Step3 Get the short sequences set of all sample sets.\
	
	Process all of the sequence sample sets according to step1 and step2, and get a short sequence set $S=(s_1, s_2, ..., s_k)$ without duplication.\
	
	Step4 Express the probability vector of the sequences with uncertain amount and unequal length.\
	$$
	P_n=\left(
	\begin{array}{cccc}
	C_1 & C_2 & ... & C_{b_1} \\
	C_1 & C_2 & ... & C_{b_2} \\
	& & ... & \\
	C_1 & C_2 & ... & C_{b_l} \\
	\end{array}
	\right)
	$$\
	$$
	{P_n}'=\left(
	\begin{array}{cccccc}
	S & C_1 & C_2 & ... & C_{b_1} & E \\
	S & C_1 & C_2 & ... & C_{b_2} & E \\
	S & & & ... & & E \\
	S & C_1 & C_2 & ... & C_{b_l} & E \\
	\end{array}
	\right)
	$$\
	
	For a sequence set $P_n$ corresponding to a certain IP, there are l function code sequences with unequal length and the lengths of them are $b_1,b_2, ... ,b_l, b_i \geq 1,i \in [1,l]$. By adding the start and the end status to each sequences, we get ${P_n}'$. And then we process all the function codes sequences with the slide window method to construct a feature vector $X_{ip}$ according to the frequency of these short sequences.\
	$$ X_{ip}=(X_{s_1},X_{s_2}, ..., X_{s_k}), \sum_{i=1}^{k}X_{s_i}=1 $$\
	The method for FCS feature vector processing is shown in figure 10. We make an improvement on the short sequence processing method in literature [13]. The improved method has the following advantages: Firstly, we transform the FCS with uncertain amount and unequal length from the same IP into feature vectors with the same length, and we retain the information of the function codes and their parameters resorting to the frequency characteristics of the short sequence. Secondly, when the length of the short sequence is set to 3, we can process the sequences with unequal length including the length of 1 or 2, by adding the start and the end status.\
	
	\begin{figure}[t]
		\begin{center}
			\includegraphics[width=0.48\textwidth]{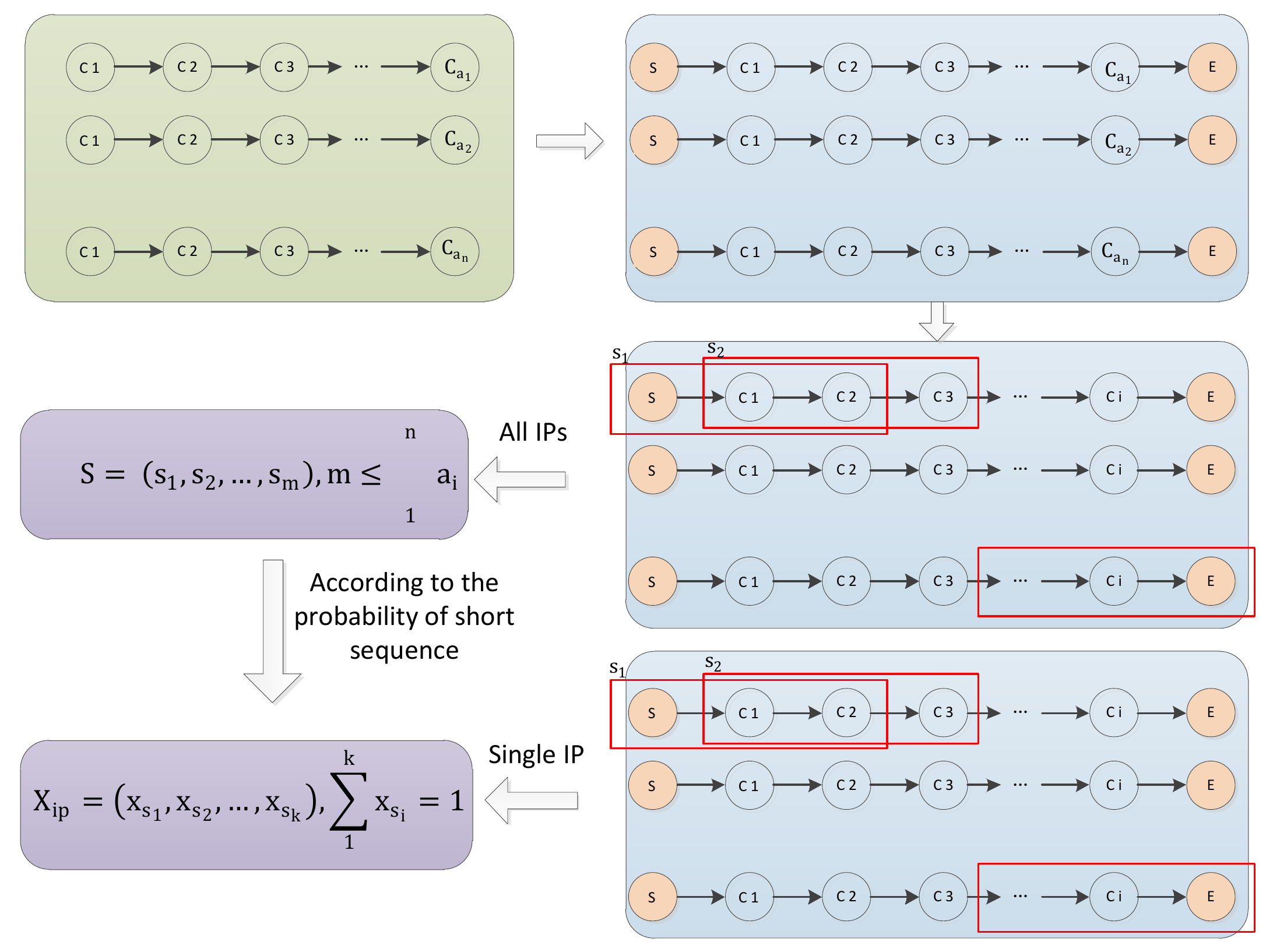}
		\end{center}
		\caption{Method for FCS feature vector processing.}
	\end{figure}
	
	Parameters sequence (PS) indicates the change rule of the parameters in all the function codes used by the sessions from the same IP and it is arrayed by chronological order. Similar to FCS, we use the same method to process PS.\\
	indicates the rule of how the parameters vary in all the function codes used by the sessions from the same IP and it is also arrayed by chronological order. Similar to FCS, we use the same method to process PS.\\
	
	\section{Partial Seeded K-Means Algorithm}
	
	We have tried machine learning methods for malicious IP traceback. Commonly used machine learning methods include decision tree, SVM and neural network, but all these methods need supervised training samples. But in the homology test of attacking data, the attack source is unknown and therefore the sample data has no labels. Unsupervised learning can reveal the inherent nature and law of data by learning the unlabeled training samples. Clustering is the most widely used method in unsupervised learning. Clustering is to divide the data samples into multiple classes or clusters, so that the samples in the same cluster have a higher degree of similarity and the samples in different clusters are more different from one another.\
	
	K-Means \cite{jain2010data} algorithm is one of the most classical clustering methods based on partition. The basic idea is to cluster around K points as centers in space, by classifying other samples which are the closest to them. The values of each cluster center are updated iteratively until the best clustering results are obtained. In application, the clustering effect of K-Means algorithm is greatly influenced by the initial center selection method.\
	
	Considering the clustering performance can be improved by using labeled samples to assist the initial center selection, Wagstaff et al. \cite{wagstaff2001constrained} proposed the COP K-Means algorithm. By constructing the two constraint sets of Must-list and Cannot-link, the samples were constrained when they were added to clusters, but the selection of the initial center point was not constrained. Basu et al. \cite{basu2002semi} proposed Seeded/Constrained K-Means algorithm. It constrained the choices of initial center through seed, and the constraint was also valid when a sample was added into a cluster. However, in this method, each cluster needs a pre-existing seed.\
	
	In the IP traceback process, it is possible to know that some IP addresses belong to a certain organization. However, it is very hard to know all the organizations in advance. That means some cluster do not have pre-existing seed. Therefore, we designed a Partial Seeded K-Means algorithm to solve this problem.\
	
	\begin{algorithm}[t]
		\caption{Partial Seeded K-Means}
		\textbf{Input:} Given a sample set $D=\left \{x_1, x_2, ..., x_m\right \}$, the clustering number k, the known clustering number l, $k\leq l$, the sample subset of known cluster partition$D'=\left \{x_1, x_2, ..., x_n\right \}$, and the sample subset of unknown cluster partition $D-D'$.\
		\begin{enumerate}
			\item Calculate the mean of the samples in each known cluster $C_i (1 \leq i \leq l)$: ${\mu}_i=\frac{1}{\left | {c_i} \right |}\sum_{x \in c_i}x$.
			\item Calculate the distance from each sample $x_j (1 \leq j \leq m-n)$ in $D - D'$ to the known mean ${\mu}_i (1 \leq i \leq l)$, and choose the largest value which equals mean distance added minimum distance as the new initial mean ${\mu}_{l+1}$，and let ${\mu}_{l+1}$ as known mean.
			\item Repeat step 2, until $k-l$ samples are chosen as the initial mean vector $\left \{{\mu}_{l+1}, {\mu}_{l+2}, ..., {\mu}_{l+k}\right \}$, make ${\mu}_i (i \leq i \leq l)$ and $\left \{{\mu}_{l+1}, {\mu}_{l+2}, ..., {\mu}_{l+k}\right \}$ to be the initial mean vector with k means.
			\item Calculate the distance $d_{ij} = {\left \|  x_j - {\mu}_i \right \|}_2$ which is from each sample $x_j (1 \leq j \leq m-n)$ in $D - D'$ to each mean vector ${\mu}_i (1 \leq i \leq k)$.
			\item Choose the cluster label for the sample $x_j$ according to nearest initial vector ${\lambda}_j=\arg\min_{i \in {1, 2, ..., k-l}}d_{ji} (1 \leq j \leq m-n)$, and add $x_j$ into corresponding cluster $C_{{\lambda}_i}=C_{{\lambda}_i}\cup \left \{x_j\right \}$.
			\item Calculate new mean vector ${\mu}_{i}^{'}=\frac{1}{\left | {c_i} \right |}\sum_{x \in c_i}x$,  if ${\mu}_{i}^{'}\neq {\mu}_i$ and update ${\mu}_i$ to ${\mu}_{i}^{'}$.
			\item Repeat step 4,5,6, until no mean vector to update.
		\end{enumerate}
		\textbf{Output:} Cluster partition $C=\left \{C_1, C_2, ..., C_k\right \}$
	\end{algorithm}
	Partial Seeded K-Means algorithm utilizes some sample subsets with known cluster partition (which is partial seed) as seed, to determine the initial center point. Considering there may be a variety of attack modes in an organization, constraints on seed is not applied while adding a sample into the clusters. That means the samples with known cluster partition may be classified into the original cluster or a new cluster during the process of clustering.
	\section{Evaluation}
	\subsection{IP Recall Rate of the Known Organizations}
	We use the IP addresses of the four known organizations to check how many IP addresses of the same organizations are recalled in the same cluster. The four curves from Figure 11 to Figure 14 show how the recall rate varies with different K values. Apparently, the IP addresses of Shodan, Censys and Beacon Labs are all grouped into the same cluster, when the cluster number K is set between 20 and 25. However, the highest recall rate of Ditecting's IP addresses is about 40\%. That means Ditecting's IP addresses are divided into different clusters and there may be multiple attack modes in the samples of Ditecting.\
	\begin{figure}[t]
		\begin{center}
			\includegraphics[width=0.48\textwidth]{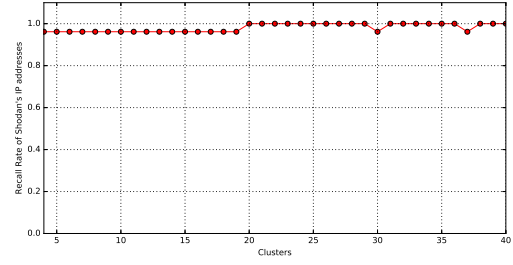}
		\end{center}
		\caption{The recall rate of Shodan's IP addresses.}
	\end{figure}
	
	\begin{figure}[t]
		\begin{center}
			\includegraphics[width=0.48\textwidth]{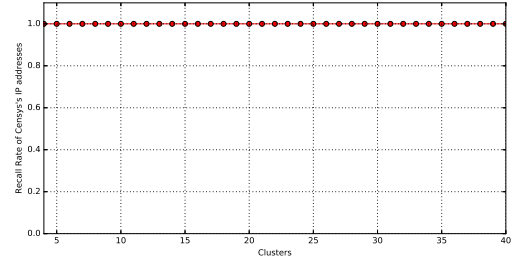}
		\end{center}
		\caption{The recall rate of Censys' IP addresses.}
	\end{figure}
	
	\begin{figure}[t]
		\begin{center}
			\includegraphics[width=0.48\textwidth]{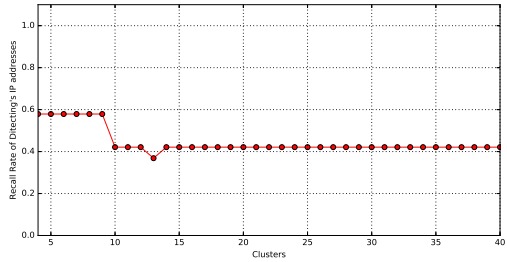}
		\end{center}
		\caption{The recall rate of Ditecting's IP addresses.}
	\end{figure}
	
	\begin{figure}[t]
		\begin{center}
			\includegraphics[width=0.48\textwidth]{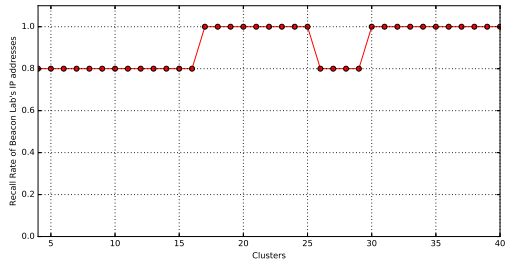}
		\end{center}
		\caption{The recall rate of Beacon Lab's IP addresses.}
	\end{figure}
	
	\subsection{Similarity Between the Predicted Value and the True Value}
	Given the knowledge of the ground truth class assignments labels\_true and our clustering algorithm assignments of the same samples labels\_pred, Adjusted Rand Index (ARI) \cite{hubert1985comparing} is a function that measures the similarity of the two assignments, ignoring permutations and with  chance normalization. Mutual Information  is a function that measures the agreement of the two assignments, ignoring permutations. Adjusted Mutual Information (AMI) is normalized against chance \cite{vinh2010information}.\
	
	We use the 66 IP addresses of the known organizations out of 573 valid IP addresses to compare the similarity between the predicted value and the true value. Figure 15 shows how ARI and AMI scores between the predicted and the true values of the 66 IP addresses vary with different K values. Apparently, the clustering works best when the number of clusters K is set between 20 and 29.
	
	\begin{figure}[t]
		\begin{center}
			\includegraphics[width=0.48\textwidth]{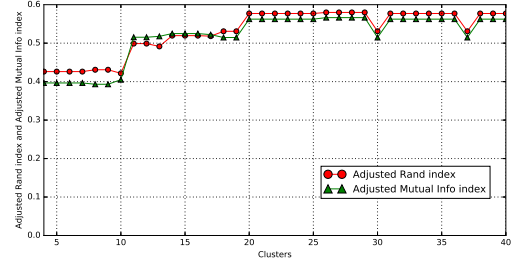}
		\end{center}
		\caption{ARI and AMI scores between the predicted and the true values of the 66 IP addresses vary with different K values.}
	\end{figure}
	
	\subsection{Clustering Performance}
	In the previous sections, we have evaluated the clustering effect using the samples with known labels. If the ground truth labels are unknown, evaluation must be performed using the model itself. The Silhouette Coefficient \cite{rousseeuw1987silhouettes} is an example of such an evaluation, where a higher Silhouette Coefficient score relates to a model with better defined clusters. Calinski-Harabaz index \cite{calinski1974dendrite} can be used to evaluate the model too, where a higher Calinski-Harabaz score relates to a model with better defined clusters.\
	
	Figure 16 and Figure 17 respectively show the curves of Silhouette Coefficient score and Calinski-Harabaz score, when the number of clusters K is set differently. Apparently, the clustering works best when K is set to 20.
	
	\begin{figure}[t]
		\begin{center}
			\includegraphics[width=0.48\textwidth]{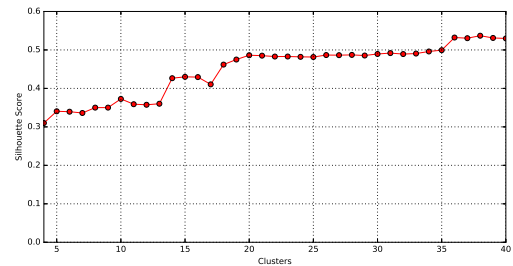}
		\end{center}
		\caption{Silhouette Coefficient score vary with different K values.}
	\end{figure}
	
	\begin{figure}[t]
		\begin{center}
			\includegraphics[width=0.48\textwidth]{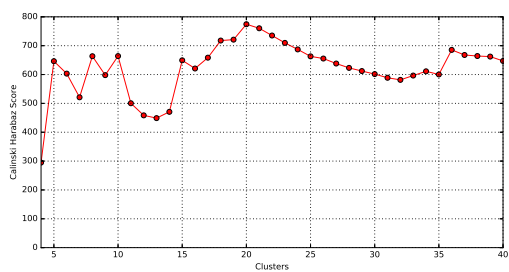}
		\end{center}
		\caption{Calinski-Harabaz score vary with different K values.}
	\end{figure}
	
	\subsection{Attack Pattern Recognition}
	Figure 18 shows the total number of clusters, in which those IP addresses of the four known organizations are grouped. No matter what value K is set, the maximum number of clusters is always 6. It indicates that there are only 6 attack patterns at the most in the samples with known organization labels.\
	
	\begin{figure}[t]
		\begin{center}
			\includegraphics[width=0.48\textwidth]{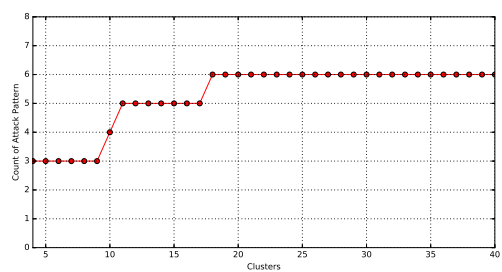}
		\end{center}
		\caption{The total number of clusters, in which those IP addresses of the four known organizations are grouped.}
	\end{figure}
	
	The attack pattern of Shodan, Censys, and Beacon Lab is unique, when the cluster number K is set between 20 and 25. But Detecting's attack mode is not unique. All the IP addresses of Detecting belong to three different clusters, except that four IP addresses are labeled as Shodan and two IP addresses are labeled as Censys. The specific distribution of these IP addresses is shown in Figure 19.
	
	\begin{figure}[t]
		\begin{center}
			\includegraphics[width=0.48\textwidth]{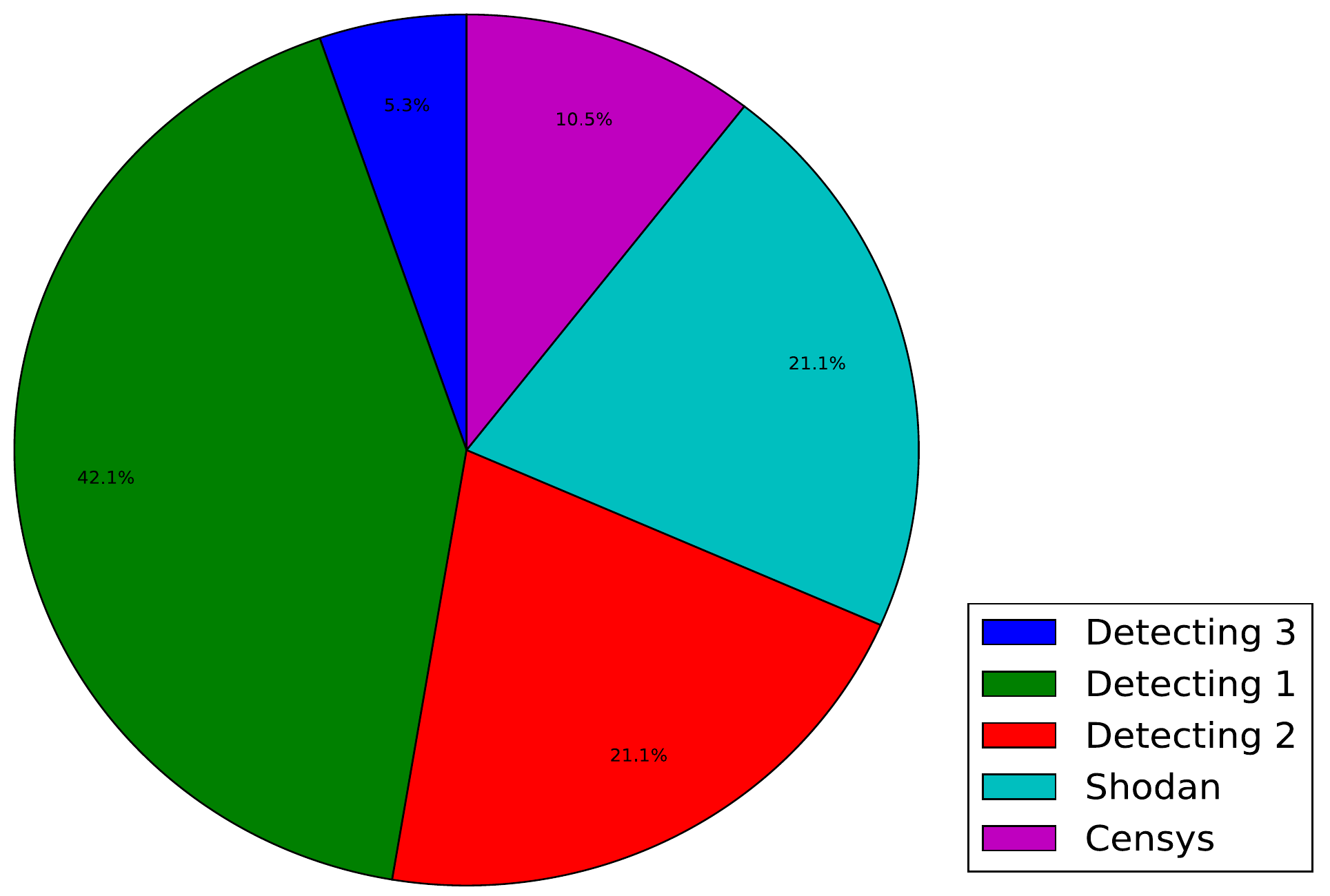}
		\end{center}
		\caption{Distribution of Detecting's IP addresses.}
	\end{figure}
	
	\subsection{Organization Identificaion}
	We set the cluster number K to be 20 for clustering and get 20 clusters at last. That means we find 20 kinds of attack patterns. However, these 20 attack patterns do not indicate there are 20 organizations. Because an organization may have multiple attack patterns, and some different organizations may also share a common attack pattern. The DNS query results and the geographical locations of IP Addresses are helpful to identify the organizations. If the IP addresses in a cluster point to the same static domain name or they are very close geographically, we can name this cluster with these labels.\
	
	\begin{table*}
		\centering
		\begin{tabular}{c*{4}c}
			\toprule
			\textbf{Cluster} & \textbf{IP Count} & \textbf{Auxiliary Information} & \textbf{Organization} \\
			\midrule
			1 & 93 & 22 IP are mapped to the domain name \textbf{shodan.io} & Shodan \\
			2 & 180 & 14 IP are mapped to the domain name \textbf{eecs.umich.edu} & Censys \\
			3 & 75 & 8 IP are mapped to the domain name \textbf{neu.edu.cn} & Ditecting \\
			4 & 43 & 5 IP are mapped to the domain name \textbf{plcscan.org} & Beacon lab \\
			11 & 51 & 26 IP are mapped to the \textbf{dynamic} domain name \textbf{binaryedge.ninja} & binaryedge.ninja \\
			13 & 11 & 4 IP are mapped to the domain name \textbf{neu.edu.cn} & Ditecting \\
			14 & 17 & 6 IP are mapped to the \textbf{dynamic} domain name \textbf{amazonaws.com} & amazonaws.com \\
			15 & 20 & 17 IP are located in \textbf{China} & China Org \\
			17 & 35 & 25 IP are mapped to the \textbf{dynamic} domain name \textbf{members.linode.com} & linode.com \\
			18 & 14 & 11 IP are located in \textbf{China} & China Org \\
			19 & 14 & 12 IP are located in\textbf{ Europe} & Europe Org \\
			20 & 9 & 7 IP are located in \textbf{China} & China Org \\
			\bottomrule
		\end{tabular}
		\caption{Clusters and their labels of organization.}
	\end{table*}\
	
	As shown in Table 5, there are 20 clusters with no less than 9 IP addresses in each of them. According to the DNS query results, Some IP addresses in cluster 1, 2, 3 and 4 point to a static domain name, and some IP addresses in the clusters 11, 14, and 17 point to a dynamic domain name. There is no domain name for reference in clusters 15, 18, 19 and 20. However, they are located in a particular country or a region, so we can name these clusters with the geographical labels. Furthermore, the cluster 3 and 13 are labeled as Ditecting, which confirms the existence of multiple attack patterns in a single organization.
	
	\section{Related Work}
	\subsection{ICS Intrusion Detection}
	Khalili and Sami \cite{khalili2015sysdetect} have proposed the SysDetect, which is a Systematic approach to Critical State Determination, to solve the problem of determining the critical states in the state-based intrusion detection. This system built a well-established and iterative data mining algorithm, ie Apriori. Kwon et al. \cite{kwon2015behavior} have proposed a novel behavior-based IDS for IEC 61850 protocol using both statistical analysis of traditional network features and specification-based metrics. Yang et al. \cite{yang2013rule} have presented a rule-based IDS for IEC 60870-5-104 driven SCADA networks using an in-depth protocol analysis and a Deep Packet Inspection (DPI) method. McParland et al. \cite{mcparland2014monitoring} have proposed the characteristic-based intrusion detection, which is an extension of the specification-based method, by defining a set of good properties and looking for behavior outside those properties. A specification-based intrusion detection model is designed to enhance the protection from both outside attacks and inside mistakes through combining the command sequence with the physical device sensor data. Mo et al. \cite{mo2014detecting} have developed the model-based techniques which is capable of detecting integrity attacks on the sensors of a control system. It is assumed that the attacker wishes to disrupt the operation of a control system in steady state, to which end the attacker hijacks the sensors, observes, and records their readings for a certain amount of time, and repeats them afterward to camouflage his attack. The model-based techniques can effectively prevent such attacks. Shang et al. \cite{shang2014modbus} have presented PSO-SVM algorithm which optimizes parameters by advanced Particle Swarm Optimization (PSO) algorithm. The method identifies anomalies of Modbus TCP traffic according to appear frequencies of the mode short sequence of Modbus function code sequence. Zhou et al. \cite{zhou2015design} have designed a novel multimodel-based anomaly intrusion detection system with embedded intelligence and resilient coordination for the field control system in industrial process automation. In this system, a multi-model anomaly detection method is proposed, and a corresponding intelligent detection algorithm is designed. In addition, in order to overcome the shortcomings of anomaly detection, a classifier based on intelligent hidden Markov model is designed to distinguish the actual attacks and failures.
	\subsection{IP Traceback}
	Savage et al. \cite{savage2000practical} have described a general purpose traceback mechanism based on probabilistic packet marking. Routers probabilistically mark packets with partial path information when they arrive. By combining a modest number of such packets, a victim can reconstruct the entire path. Snoeren et al. \cite{snoeren2001hash} have presented a hash-based technique for IP traceback that generates audit trails for traffic within the network, and can trace the origin of a single IP packet delivered by the network in the recent past. Belenky et al. \cite{belenky2003ip} have proposed a Deterministic Packet Marking algorithm, which only requires the border router to mark the 16-bits Packet ID field and the reserved 1-bit Flag in the IP header. Therefore, the victim can obtain the corresponding entry address and the subnet where the attack source is located. This method is simple and efficient compared to Probabilistic Packet Marking algorithm. Bellovin et al. \cite{bellovin2003icmp} have proposed an ICMP Traceback Message. When forwarding packets, routers can, with a low probability, generate a traceback message that is sent along to the destination or back to the source. With enough traceback messages from enough routers along the path, the traffic source and path of forged packets can be determined. Goodrich et al. \cite{goodrich2008probabilistic} have presented a new approach to IP traceback based on the probabilistic packet marking paradigm. This approach, which is called randomize-and-link, uses large checksum cords to link message fragments in a way that is highly scalable, for the cords serve both as associative addresses and data integrity verifiers. The main advantage of this approach is that attacker cannot fabricate a message and it has good scalability. Gong et al. \cite{gong2005ip,gong2008more} have presented a novel hybrid IP traceback approach based on both packet logging and packet marking. They maintain the single packet traceback ability of the hash-based approach and, at the same time, alleviate the storage overhead and access time requirement for recording packet digests at routers. Their work improves the practicability of single-packet IP traceback by decreasing its overhead. Yang et al. \cite{yang2012riht} have proposed a traceback scheme that marks routers’ interface numbers and integrates packet logging with a hash table (RIHT) to deal with the logging and marking issues in IP traceback. RIHT has the properties of low storage, high efficiency, zero false positive and zero false negative rates in attack-path reconstruction. Yu et al. \cite{yu2016feasible} have proposed a marking on demand (MOD) scheme based on the DPM mechanism to dynamically assign marking IDs to DDoS attack related routers to perform the traceback task. They set up a global mark distribution server (MOD server) and some local DDoS attack detector. When there appears suspicious network flows, the detector requests unique IDs from the MOD server, and embeds the assigned unique IDs to mark the suspicious flows. At the same time, the MOD server deposits the IP address of the request router and the assigned marks, which are used to identify the IP addresses of the attack sources respectively，into its MOD database. Fadel et al. \cite{fadel2016low} have presented a new hybrid IP traceback framework. This framework is based on both marking and logging techniques. In the marking algorithm, every router is assigned a 12-bits-length ID number; it helps in deploying pushback method to permit legitimate traffic flow smoothly. In the packet logging technique, a logging ratio is managed by changing a value k specified in the traceback system. This framework can save more than 50\% of the storage space of routers. Cheng et al. \cite{cheng2017fact} argue that cloud services offer better options for the practical deployment of an IP traceback system. They have presented a novel cloud-based traceback architecture, which possesses several favorable properties encouraging ISPs to deploy traceback services on their networks. This architecture includes a temporal token-based authentication framework, called FACT, for authenticating traceback service queries. Nur et al. \cite{nur2018record} exploit the record route feature of the IP protocol, and propose a novel probabilistic packet marking scheme to infer forward paths from attacker sites to a victim site and enable the victim to delegate the defense to the upstream Internet Service Providers (ISPs). Compared to the other techniques, this approach requires less many packets to construct the paths from attacker sites toward a victim site.
	\section{Conclusions}
	IP traceback for cyber attacks usually needs redesigning the Internet deploying new service. In this study, we have proposed a malicious IP traceback model, i.e. ICSTrace, for Industrial Control System without changing the Internet infrastructure or deploying any new services. By analyzing the characteristics of the attack data, we extract the numeric features and the sequence transformation features from the function codes and their parameters. Those features are expressed by a one-dimensional vector, which stands for the unique pattern of an attack. As a result, the problem of IP traceback turns into a problem of clustering those patterns. We also propose a Partial Seeded K-Means algorithm to cluster the IP addresses with the same pattern into a malicious organization. The effectiveness of ICSTrace is proved by experiments on real attack data. Although ICSTrace can not recover the whole path of the attack, it is significant in the following aspects:\
	\begin{enumerate}
		\item Find out the malicious IP addresses which belong to the same organization.
		\item Reveal the unexposed active IP addresses belonging to the known organizations.
		\item Collect the springboards used by the same organization for launching attacks.
		\item Provide learning samples for subsequent malicious behavior identification by expressing the attack pattern in the form of feature vector.
	\end{enumerate}
	
	\section{Future work}
	In the future, we will improve ICSTrace and apply it to other kinds of ICS protocols, even the traditional Internet protocols. At the same time, we will use the attack patterns as the learning samples to design and validate the intrusion detection system based on machine learning, to solve the difficult problem of unknown threat detection.\
	
	\section{Acknowledgments}
	The authors thank Biao Chang, Binglei Wang and Dazhong Shen for their useful feedback and comments.\
	
	{\footnotesize \bibliographystyle{acm}
		\bibliography{bibliography}
	}
	
\end{document}